\documentclass[12pt]{article}
\usepackage{amsfonts,amssymb,graphicx}

\title{Opposite Arrows of Time Can Reconcile Relativity and
Nonlocality}
\author{Sheldon Goldstein\footnote{Department of Mathematics, Rutgers
  University, New Brunswick, NJ 08903, USA}\ \ and Roderich
  Tumulka\footnote{Mathematisches Institut der Universit\"{a}t
  M\"{u}nchen, Theresienstra{\ss}e 39, 80333 M\"{u}nchen, Germany}}
\date{}

\newcommand{\CCC}{\mathbb{C}}
\newcommand{\RRR}{\mathbb{R}} %
\newcommand{\1}{\mathbf{1}} %
\newcommand{\st}{(\mbox{space-time})}
\newcommand{\fC}{\mbox{future}_C}
\newcommand{\pC}{\mbox{past}_C}
\newcommand{\pTh}{\mbox{past}_\Theta}
\newcommand{\fTh}{\mbox{future}_\Theta}
\newcommand{\ret}{\mathrm{ret}}
\newcommand{\const}{\mathrm{const}}

\begin{document}
\maketitle

\begin{abstract}
We present a quantum model for the motion of $N$ point particles, implying
nonlocal (i.e., superluminal) influences of external fields on the
trajectories, that is nonetheless fully relativistic. In contrast to other
models that have been proposed, this one involves no additional space-time
structure as would be provided by a (possibly dynamical) foliation of
space-time. This is achieved through the interplay of opposite microcausal
and macrocausal (i.e., thermodynamic) arrows of time.\\ PACS numbers
03.65.Ud; 03.65.Ta; 03.30.+p
\end{abstract}
%\openup-.1\jot

\section{Introduction}

We challenge in this paper a conclusion that is almost universally
accepted: that quantum phenomena, relativity, and realism are
incompatible. We show that, just as in the case of the no-hidden-variables
theorems, this conclusion is hasty.  And, as in the hidden variables
case, we do so with a counterexample.

We present a relativistic toy model for nonlocal quantum phenomena that
avoids the usual quantum subjectivity, or fundamental appeal to an
observer, and describes instead, in a rather natural way, an objective
motion of particles in Minkowski space. In contrast to that of \cite{HBD},
see below, our model invokes only the structure at hand: relativistic
structure provided by the Lorentz metric and quantum structure provided by
a wave function. It shares the conceptual framework---and forms a natural
generalization---of Bohmian mechanics, a realistic quantum theory that
accounts for all nonrelativistic quantum phenomena \cite{BM}.  The key
ingredient is a mechanism for a kind of mild backwards causation, allowing
only a very special sort of advanced effects, that is provably
paradox-free.

Unfortunately, the model considered here, unlike that of \cite{HBD},
does not provide any obvious, distinguished probability measure on the
set of its possible particle paths, on which many of its detailed
predictions are likely to be based. It is thus difficult to assess the
extent to which the model is consistent with violations of Bell's
inequality \cite{Bell}. However, in the nonrelativistic limit of small
velocities (and slow changes in the wave function), the behavior of
our model coincides with that of the usual Bohm--Dirac model \cite{BD}
and is thus consistent with the $|\psi|^2$ distribution, and hence
with violations of Bell's inequality, in every frame in which the
velocities are slow.

\section{Spirit of the Model}

The backwards causation arises from a time-asymmetric equation of motion
for $N$ particles that involves advanced data about the other particles'
world lines.  The asymmetry of this law defines an intrinsic arrow of time,
which is not present in well-known theories like Newtonian mechanics or
Wheeler--Feynman electrodynamics, and which we call the \emph{microcausal
arrow of time}, as opposed (and, indeed, opposite) to the usual,
thermodynamic or \emph{macrocausal arrow of time}.

In a recent paper \cite{Schulman}, L.~S.~Schulman investigated the
possibility of opposite thermodynamic arrows of time in different regions
of the universe: that in some distant galaxy, entropy might \emph{decrease}
with (our) time, ``eggs uncrack,'' and inhabitants, if present, feel the
arrow of time to be just opposite to what we feel. He studied this question
in terms of statistical mechanics, and, on the ground of computer
simulations, came to the conclusion that this is quite possible, apparent
causal paradoxes notwithstanding. We also consider two opposite arrows 
of time, but not belonging to different regions of space-time, and 
not as a study in statistical mechanics, but as a possible explanation 
of quantum nonlocality. Instead of having the thermodynamic arrow
of time vary within one universe, we consider the situation in which two
conceptually different arrows of time, the microcausal and the
macrocausal arrow, are everywhere opposite throughout
the entire universe. 

It has been suggested \cite{HBD} that in order to account for quantum
nonlocality, one employ---contrary to the spirit of relativity---a
time-foliation, i.e., a foliation of space-time into 3-dimensional
spacelike hypersurfaces, which serve to define a temporal order for
spacelike separated points, or one might say simultaneity-at-a-distance,
and hence simultaneity surfaces along which nonlocal effects
propagate. This foliation is intended to be understood, not as a gauge
(i.e., as one among many points of view a physicist may choose), but as an
additional element of space-time structure existing objectively out there
in the universe, defining in effect a notion of true simultaneity. In
\cite{HBD}, the time-foliation is itself a dynamical variable subject to an
evolution law. In contrast, the model we present here does not invoke a
distinguished foliation.

In our model, the formula for the velocity of a particle at space-time
point $p$ involves, in a Lorentz-invariant manner, the points where the
world lines of the other particles intersect---not any ``simultaneity
surface'' containing $p$ but rather---the future light cone of $p$,
as well as the velocities of the particles at these points. As a
consequence, it is easy to compute the past world lines from the future
world lines, but it is not at all obvious how to compute the future from
the past---except by testing all the uncountably many possibilities. One
can say that the behavior of a particle at time $t$ has causes that lie in
the future of $t$, so that on the microscopic level of individual particles
and their world lines, the arrow of time of causation, as defined by the
dynamics, points towards the past. We call this the \emph{microcausal arrow
of time} and denote it by $C$; it defines a notion of ``$\fC$'' = past, and
of ``$\pC$'' = future. Thus the velocity of a particle depends on where the
other particles intersect the $\pC$ light cone of that particle (effects
are ``retarded$_C$''). This microcausal arrow of time is also an arrow of
determinism: knowledge of the world lines prior$_C$ to a certain time
determines the $\fC$, whereas there is no reason to believe the converse,
that the $\fC$ determines the $\pC$.

\begin{figure}[h]
\begin{center}
\includegraphics[width=.3\textwidth]{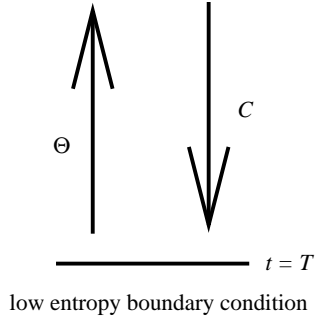}
\end{center}
\caption[]{Boundary conditions are imposed on the $\fC=\pTh$ end of
time. (The time direction is vertical.)}
\label{arrowf2}\end{figure}

Now consider the set of solutions of the law of motion as given, and
consider those solutions which at a certain time $T$ in the distant $\fC$
reside in a certain macrostate with low entropy. We are interested in their
behavior for times prior$_C$ to $T$. One should expect that entropy
decreases in the direction of $C$ until it reaches its minimum at $T$. So
the thermodynamic arrow of time $\Theta$, as defined by the direction of
entropy increase, is opposite to $C$ (see Fig.~\ref{arrowf2}).

The arrow of time that inhabitants of this imaginary world would perceive
as natural is the one corresponding to eggs cracking rather than
uncracking, that is, the thermodynamic one. So when the inhabitants speak
of the future, they mean $\fTh = \pC$. That is why we called $\pC$ the
future in the beginning. It is $\Theta$ that corresponds to macroscopic
causality.

\begin{figure}[h]
\begin{center}
\includegraphics[width=.4\textwidth]{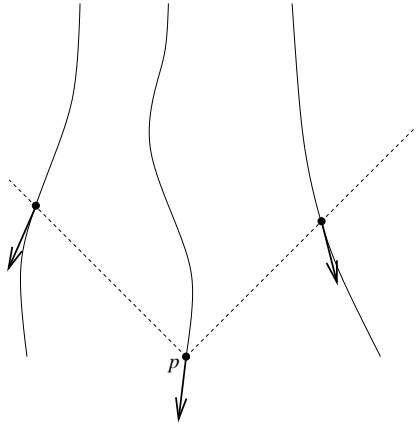}
\end{center}
\caption[]{
%For computing the 4-velocity of a particle at a space-time point
%$p$, information about where the world lines of the other particles cross
%the $\pC$ light cone of $p$, and the 4-velocities at these points, is
%relevant.
The 4-velocity of a particle at a space-time point $p$ depends on
where the world lines of the other particles cross the $\pC$ light
cone of $p$, and on the 4-velocities at these points.
}
\label{arrowf4}\end{figure}

The law of motion is Lorentz-invariant, and since it involves
retarded$_C$ but not advanced$_C$ influences, it is local with respect
to $C$, i.e.\ what is happening at a space-time point $p$ depends only
on what happened within (and on) the $\pC$ light cone of $p$. With
respect to $\Theta$, however, the law of motion is \emph{nonlocal}, as
the trajectory of a particle at $p$ depends on where and how the other
particles cross the $\fTh$ light cone of $p$ (see Fig.~\ref{arrowf4}),
which again might be influenced by interventions of macroscopic
experimenters at spacelike separation from $p$. So this law of motion
provides an example of a theory that entails nonlocality (superluminal
influences) while remaining fully Lorentz-invariant.

It is important here to appreciate that the thermodynamic arrow of
time arises not from any microscopic time asymmetry, but from boundary
conditions of the universe. That is a moral of Boltzmann's analysis of
the emergence of the thermodynamic arrow of time from a time symmetric
microscopic dynamics, where no microscopic time arrow is available as
a basis of macroscopic asymmetry. As a consequence, the microscopic
asymmetry present in our model should not affect the thermodynamic
arrow $\Theta$ at all, and we should be free to choose the direction
of $\Theta$ as either the same as or opposite to $C$, by imposing
low-entropy boundary conditions at either the distant past$_C$ or
future$_C$ when setting up the model. The advantage of having them
opposite is that this allows our model to display \emph{nonlocal}
behavior. Had we chosen $\Theta$ to be in the same direction as $C$,
then the model would have been \emph{local}---because then what
happens at $p$ would depend only on what had happened in the past
light cone of $p$.

\section{Equations of the Model}

Now let us turn to the details of the model. It is similar to Bohmian
mechanics \cite{BM}, in the sense that velocities are determined by a
wave function. In our case, the wave function is an $N$ particle Dirac
spinor field, i.e., a mapping $\psi:\st^N \to (\CCC^4)^{\otimes
N}$. We consider entanglement, but without interaction. The wave
function is supposed to be a solution of the multi-time Dirac equation
$$
  \1\otimes\cdots\otimes\underbrace{\gamma^\mu}_{i\mathrm{th\:\: place}}  
  \otimes\cdots\otimes\1 \:
  (i\hbar\partial_{i,\mu}+eA_\mu(x_i))\: \psi = m\psi
$$
where summation is understood for $\mu$ but not for $i$, $m$ is the mass
parameter, $e$ the charge parameter, $\gamma^\mu$ are the Dirac matrices,
$A_\mu$ is an arbitrary given 1-form (the external electromagnetic vector
potential), and $i$ runs from 1 through $N$ enumerating the particles. 

For any space-time point $p$ and any parametrized timelike curve
$x^\mu(s)$, let $s_\ret(p)$ denote the value\footnote{One might worry about
the existence and uniqueness of these points, and rightly so: whereas
uniqueness is a consequence of the world line's being timelike, existence
is actually not guaranteed. A counterexample is $x^\mu(t) =
(t,0,0,\sqrt{1+t^2})$. But we will ignore this problem here.} of $s$ such
that $x^\mu(s)$ lies on the $\pC$ light cone of $p$. Our law of motion
demands of the world lines $x^\mu_i(s_i)$ of the $N$ particles that (a)
they be timelike and (b) for every particle $i$ and parameter value $s_i$,
\begin{equation}\label{eqmotion}
  \frac{dx^{\mu_i}_i}{ds_i} \: \bigg\| \: \overline{\psi} \,
  (\gamma^{\mu_1}\otimes \cdots\otimes\gamma^{\mu_N})\, \psi
  \prod_{j\neq i} \frac{dx^{\nu_j}_j}{ds_j}\big(s_{\ret,j}(p_i)\big)
  \eta_{\mu_j\nu_j}\,,
\end{equation}
where $\|$ means ``is parallel to'' (i.e.\ is a multiple of),
$p_i=x_i(s_i)$, $s_{\ret,j}$ refers to the $x_j$ world line,
$\overline{\psi}=\psi^{\dagger}\gamma^0 \otimes \cdots \otimes
\gamma^0$, $\eta= \mbox{diag}(1,-1,-1,-1)$ is the Minkowski metric,
and $\psi$ and $\overline{\psi}$ are evaluated at $\left( x_1 \left(
s_{\ret,1}(p_i) \right),\ldots, x_N \left( s_{\ret,N}(p_i) \right)
\right)$.

Here is what the law says. Suppose that space-time point $p_i$ is on
the world line of particle $i$. Then the velocity of particle $i$ at
$p_i$ is given as follows: Find the points of intersection $p_j$ of
the world lines of the other particles with the $\pC$ light cone of
$p_i$, and let $u^\nu_j=dx^{\nu}_j/ds_j$ be the 4-velocities at these
points.  (It does not matter whether or not $u$ is normalized, $u^\nu
u_\nu=1$.) Evaluate the wave function at $(p_1,\ldots,p_N)$ to obtain
an element $\psi$ of $(\CCC^4)^{\otimes N}$, and compute also
$\overline{\psi}$. Use these to form the tensor $J^{\mu_1\ldots\mu_N}
= \overline{\psi} \gamma^{\mu_1} \otimes \cdots \otimes \gamma^{\mu_N}
\psi$. $J$ is an element of $T_{p_1}M \otimes \cdots \otimes T_{p_N}
M$, where $M$ stands for the space-time manifold and $T_p$ denotes the
tangent space\footnote{Since the space-time manifold is simply
Minkowski space, all the tangent spaces are isomorphic in a canonical
way. Nevertheless it might be helpful for didactical reasons to
distinguish between different tangent spaces.} at the point $p$. Now
for all $j\neq i$, transvect $J$ with $u^{\nu_j}_j
\eta_{\mu_j\nu_j}$. This yields an element $j_i^{\mu_i}$ of $T_{p_i}
M$, defining a 1-dimensional subspace $\RRR j_i^{\mu_i}$ of $T_{p_i}
M$. The world line of particle $i$ must be tangent to that
subspace. (One easily checks that this prescription is purely
geometrical: it provides a condition on the collection of space-time
paths that does not depend on how they are parametrized. The velocity
is not defined if $\psi( p_1, \ldots, p_N) = 0$---and only in that
case, as we will see below.)

How does one arrive at this law? To begin with, there is an obvious
extension of Bohmian mechanics to a single Dirac particle
\cite{BD}. Let $\psi:\st \to \CCC^4$ obey the Dirac equation, let
$j^\mu = \overline\psi \, \gamma^\mu \, \psi$ be the usual Dirac
probability current, and let the integral curves of the 4-vector field
$j^\mu$ be the possible world lines of the particle, one of which is
chosen at random. The world lines are timelike, and therefore
intersect every spacelike hyperplane precisely once. If this point of
intersection is $|\psi|^2$ distributed in one frame at one time, it is
$|\psi|^2$ distributed in every frame at every time. That is because
the Dirac equation implies that $j^\mu$ is a conserved current,
$\partial_\mu j^\mu =0$, and because in every frame $|\psi|^2=j^0$.
For $N=1$, our law of motion reproduces this single-particle law.

This Bohm--Dirac law of motion possesses an immediate many-particle
analogue if one is willing to dispense with covariance \cite{BD}:
using the simplest tensor quadratic in $\psi$, $J^{\mu_1 \ldots
\mu_N}= \overline\psi \, \gamma^{\mu_1} \otimes \cdots \otimes
\gamma^{\mu_N} \, \psi$, one can write
\begin{equation}\label{BD}
  \frac{dx^{\mu_i}_i(s)}{ds} \: \bigg\| \: J^{0\ldots \mu_i \ldots 0}
  (p_1,\ldots,p_N)
\end{equation}
for the velocities, where $p_1,\ldots,p_N$ are simultaneous (with
respect to one preferred Lorentz frame) and $s$ is again an arbitrary
curve parameter. This motion also conserves the probability density
$\rho = |\psi|^2 = J^{0\ldots 0}$. It can be generalized \cite{HBD} to
arbitrary spacelike hypersurfaces (rather than the parallel
hyperplanes corresponding to a Lorentz frame). The hypersurfaces then
play a twofold role: first, the multi-time field $J$ is evaluated at
$N$ space-time points $p_1,\ldots,p_N$ which are taken to lie on the
same hypersurface. Second, the unit normal vectors on the hypersurface
are used for contracting all but one of the indices of $J^{\mu_1
\ldots \mu_N}$ to arrive at a 4-vector \cite{HBD}; that is how the
0-components arise in (\ref{BD}). So (\ref{eqmotion}) is merely a
modification of known ``Bohmian'' equations, using a simple strategy
for avoiding the use of distinguished spacelike hypersurfaces: use
light cones as the hypersurfaces for determining the $N$ space-time
points, and use the velocity 4-vectors (of the other particles) for
contracting all but one of the indices of $J$. (If vectors normal to
the light cone had been used, the model would not have a good
nonrelativistic limit; see below.)

``Bohmian'' equations of motion usually imply that positions can be
taken to always be $|\psi|^2$ distributed. That is what makes Bohmian
mechanics compatible with the empirical facts of quantum mechanics. In
contrast, our velocity formula (\ref{eqmotion}) does not conserve the
$|\psi|^2$ distribution, and that is why we call it a toy model rather
than a serious theory. In the nonrelativistic limit, however, our
model coincides with the many-particle Bohm--Dirac law, since the
future light cone approaches the $t=\const$ hyperplane, and hence is
compatible with a $|\psi|^2$ distribution, consistent with quantum
mechanics. Note also that, due to Bell's theorem \cite{Bell}, a
necessary condition for compatibility of a law of motion with the
$|\psi|^2$ distribution is its nonlocality. So a necessary step
towards such a law that is relativistic is to come up with a covariant
method of providing nonlocality: we develop one such method here.

The question remains as to whether for $\psi\neq 0$, $j_i^{\mu}$ is
timelike. Actually, it is sometimes lightlike: e.g., for
$\psi(p_1,\ldots,p_N) = \frac{1}{2}(1,1,1,-1) \otimes \psi'$ in the
standard representation with $\psi' \in (\CCC^4)^{\otimes(N-1)}$, one finds
that $j_1^\mu = (1,0,0,1)$. But this is an exceptional case like $\psi=0$. To
see that $j_i^{\mu}$ is either timelike or lightlike, note that a vector is
nonzero-timelike-or-lightlike if and only if its scalar product with every
timelike vector is nonzero. So pick a nonzero timelike vector, call it
$u_i^{\nu_i}$, and compute the scalar product 
$$
  \lambda := j_i^{\mu_i}
  u_i^{\nu_i} \eta_{\mu_i\nu_i} 
  = \overline{\psi} \, (\gamma^{\mu_1}\otimes
  \cdots\otimes\gamma^{\mu_N})\, \psi \: \prod_{j} u_j^{\nu_j}
  \eta_{\mu_j\nu_j}\,. 
$$
Without changing the absolute value of $\lambda$, we
can make sure all $u_j$s are future-pointing (i.e.\ $u_j^0>0$), replacing
$u_j$ by $-u_j$ if necessary. Through a suitable choice of $N$ Lorentz
transformations in the spaces $T_{p_1}M,\ldots,T_{p_N}M$, we can replace
all $u_j$s by $(1,0,0,0)$ while also replacing $\psi$ by a transformed
spinor $\psi'$. Thus $\pm \lambda = \overline{\psi'} \, (\gamma^{0}\otimes
\cdots\otimes\gamma^{0})\, \psi'= (\psi')^\dagger \psi'>0$ unless
$\psi'=0$, i.e., unless  $\psi=0$.

\section{Properties of the Model}

\begin{figure}[h]
\begin{center}
\includegraphics[width=.8\textwidth]{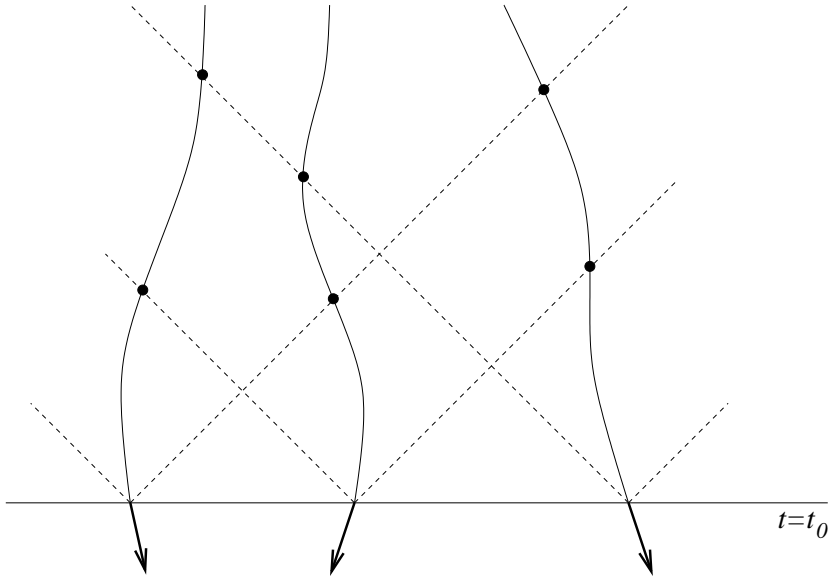}
\end{center}
\caption[]{
%For computing velocities (bold arrows) at $t=t_0$, information
%is relevant about where (dots) the world lines (curves) cross the $\pC$
%light cones (dashed lines), and what the velocities are at these points.
For computing velocities at $t=t_0$, information about several space-time 
points, lying at different times, is relevant: where the crossing points 
(dots) through the light cones are, and what the velocities are at these 
points.
}
\label{arrowf1}\end{figure}

The model is of course very restricted in the sense that we do not allow
for interaction between the particles. While it is difficult to find global
solutions to this law of motion, it is quite obvious how to obtain a
solution from initial$_C$ (=final$_\Theta$) boundary counditions: for
computing the velocities of all particles at any time $t=t_0$, data are
needed about velocities of the particles at several earlier$_C$
(=later$_\Theta$) instants of time (see Fig.~\ref{arrowf1}), and all such
data are available given the world lines prior$_C$ to (=after$_\Theta$)
$t_0$. While propagation in our model from microcausal past to microcausal
future is far from routine, it is thus ordinary enough to make it seem
reasonable that there should exist a unique continuation of a given
past$_C$ to the future$_C$ that obeys the law in that future, even if the
specified past$_C$ does not. This of course does not prove, even
heuristically, the existence of global solutions, existing for all times,
past, present, and future, but for our purposes this does not matter so
much. For our purposes, i.e., for arguing for the existence of relevant
solutions, it is sufficient to consider the specification of the past$_C$
up to a certain time in the spirit of an initial (or final) boundary
condition, from which evolution takes place.

We note that our approach has little if any overlap with the proposals
of Huw Price \cite{price}, who argues that backwards causation
can ``solve the puzzles of quantum mechanics.'' Whereas Price seeks to
exploit backwards causation to \emph{avoid} nonlocality, we use it to
\emph{achieve} nonlocality in a Lorentz invariant way. Moreover, while 
our model involves advanced effects on
particle trajectories, we do not propose any advanced effects on the wave
function, as does Price \cite[p.132]{price}.

We have reason to believe that in our model, (macro) causes precede
(macro) effects. To be sure, why causality proceeds in one direction
alone is not easy to understand, even without a micro arrow.  The
usual understanding grounds this in low entropy ``initial''
conditions, and the same considerations should apply even when there
is a micro arrow, even when this arrow points (in a reasonable sense)
in the opposite direction. Consequently, macrocausality, which is what
usual causal reasoning involves, should follow the thermodynamic arrow
of time.
\begin{figure}[h]
\begin{center}
\includegraphics[width=.8\textwidth]{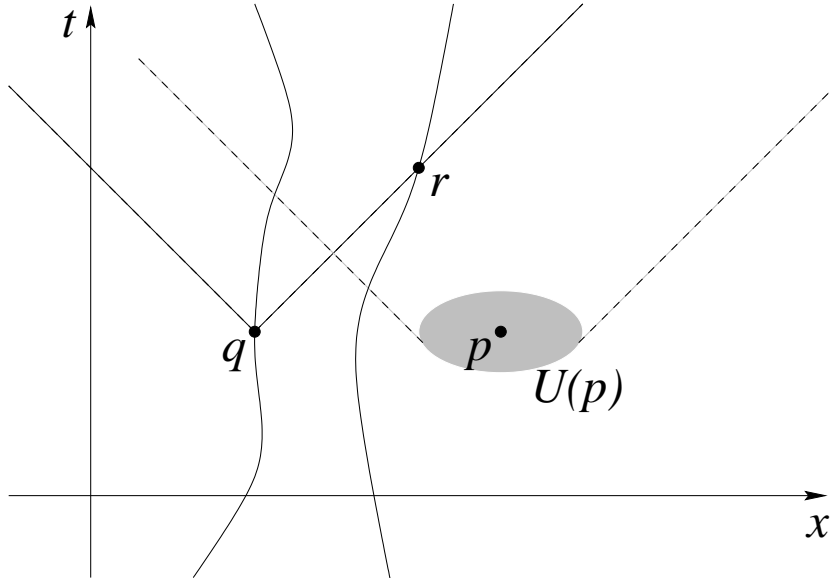}
\end{center}
\caption[]{Changing the external potential in the space-time region $U(p)$
affects the wave function only in the absolute future$_\Theta$ of
$U(p)$. But this means that $\psi(q,r)$ is changed, so that the velocity at
$q$ is likely to be affected.}
\label{Fig3}\end{figure}

To see how this interplay between micro- and macro-causality plays out in
our model, consider two electromagnetic potentials $A_\mu$ and $A_\mu'$
that differ only in a small space-time region $U(p)$ around the point
$p$ (see Fig.~\ref{Fig3}). Solving the multi-time Dirac equation for the 
same initial$_\Theta$
wave function gives two functions $\psi, \psi'$ which differ only for those
$N$-tuples $(p_1,\ldots,p_N)$ of space-time points for which at least one
$p_i$ lies inside or on the $\fTh$ light cone of some 
$\widetilde{p}\in U(p)$.  We can
regard effects on the wave function as always after$_\Theta$ the external
cause. Not so for the world lines; in general, effects of $A_\mu'(p)$ will
be found everywhere: in the future, past and present of $p$. More
precisely, given a solution $S$ (an $N$-tuple of paths) of the law of
motion for $\psi$, there will be no corresponding solution $S'$ of the law
of motion for $\psi'$ such that $S$ and $S'$ agree in the past of $U(p)$,
or in the future of $U(p)$. So on the particle level, causation is
effectively in both time directions. Note that this cannot possibly lead to
causal paradoxes since there is no room for paradoxes in the Dirac equation
and our law of motion.

This is essentially because there is no feedback mechanism which could
lead to causal loops. Instead, the Dirac equation may be solved in the
ordinary way from past to future, starting out from an
initial$_\Theta$ wave function, and the equation of motion can
subsequently be solved from initial$_C$ conditions. Thus, insofar as
the microscopic dynamics is concerned, there is no way a paradox could
possibly arise. This proves consistency even for a universe governed
as a whole by our law of motion. In this respect, the situation is
much different with Wheeler--Feynman electrodynamics, tachyons, and
theories involving closed timelike curves, since they all have
microcausal feedback, so that even the existence of solutions is
dubious for given initial data.

One last remark concerning the question, touched upon earlier, of
whether macroscopic backwards causation is possible in our model
(i.e., whether observable events in the future can cause observable
effects in the past): the nonlocal backwards microcausal mechanism in
our model is based on quantum entanglement, which is now widely
regarded as giving rise to a rather fragile sort of nonlocality,
revealed through violations of Bell's inequality, that does not
support the sort of causal relationship between observable events that
can be used for signalling. (For our model, however, things are
trickier than usual since no-signalling results are grounded on the
$|\psi|^2$ distribution.)

In the nonrelativistic limit $c\to\infty$, the unusual causal
mechanism of our model is replaced by a more conventional one: as
mentioned earlier, the future light cone (and the past light cone, as
well) approaches the $t=t_0$ hypersurface, so instead of having to
find the points where the other world lines cross the future light
cone, one needs to know the points where the other world lines cross
the $t=t_0$ hypersurface (which in the nonrelativistic limit does not
depend on the choice of reference frame). This means that the
configuration of the particle system at time $t_0$ determines directly
all the velocities and thus the evolution of the configuration into
the future (or the past, as well). Thus the nonrelativistic limit of
our model is causally routine, and our model can be regarded as
illustrating how a simple small deviation from this normal picture,
imperceptible in the nonrelativistic domain, can provide a
relativistic account of nonlocality. In fact, it is easy to see that
in the nonrelativistic limit of small velocities and slow changes in
the wave function, our model coincides with the usual Bohm--Dirac
model \cite{BD}, so that it is then compatible with a $|\psi|^2$
distribution of positions and thus with the Bell--EPRB
correlations.\footnote{However, in this limit it takes the correlated
particles much longer to pass the Stern--Gerlach magnets than $c^{-1}$
times the relevant distance, so that the results of Bell correlation
experiments that probe superluminal nonlocality are not accounted for
by our model.}

\section{Conclusions}

We have proposed that in our relativistic universe quantum nonlocality
originates in a microcausal arrow of time opposite to the thermodynamic
one. We recognize that this proposal is rather speculative. However, we
believe it is a possibility worth considering.

\bigskip

\textit{Acknowledgements.} We are grateful for the hospitality of the
Institut des Hautes \'Etudes Scientifiques (I.H.E.S.),
Bures-sur-Yvette, France, where the idea for this paper was
conceived. The paper has profited from questions raised by the
referees.

\end{document}